\documentclass[aps,prc,twocolumn,superscriptaddress,showpacs]{revtex4}
\usepackage{epsfig}

\begin{document}

\def\GSI{Gesellschaft f{\"u}r Schwerionenforschung mbH, D-64291 Darmstadt,
Germany}
\def\ROSS{Forschungszentrum Rossendorf, D-01314 Dresden, Germany}
\def\IKF{Institut f{\"u}r Kernphysik,
Universit{\"a}t Frankfurt, D-60486 Frankfurt, Germany}
\def\MIL{Istituto di Scienze Fisiche, Universit\`{a} degli Studi 
di Milano and I.N.F.N., I-20133 Milano, Italy}
\def\MSU{Department of Physics and
Astronomy and National Superconducting Cyclotron Laboratory,
Michigan State University, East Lansing, MI 48824, USA}
\def\CAT{Dipartimento di Fisica dell' Universit\`{a}
and I.N.F.N., I-95129 Catania, Italy}
\def\MAINZ{Institut f{\"u}r Kernphysik, Universit{\"a}t Mainz, D-55099 Mainz, 
Germany}
\def\SINS{Soltan Institute for Nuclear Studies,
00-681 Warsaw, Hoza 69, Poland}
\def\LANL{Subatomic Physics Group, Los Alamos National Laboratory,
Los Alamos, NM 87545, USA}
\def\KIP{Kirchhoff-Institut f{\"u}r Physik, Universit{\"a}t Heidelberg,
D-69120 Heidelberg, Germany}
\def\IFJ{H. Niewodnicza{\'n}ski Institute of Nuclear Physics, Pl-31342 Krak{\'o}w,
Poland}

\title{
Thermal and chemical freeze-out in spectator fragmentation}

\affiliation{\GSI}
\affiliation{\MIL}
\affiliation{\MSU}
\affiliation{\CAT}
\affiliation{\ROSS}
\affiliation{\IKF}
\affiliation{\SINS}
\affiliation{\LANL}
\affiliation{\KIP}
\affiliation{\IFJ}
\affiliation{\MAINZ}

\author{W.~Trautmann}		\affiliation{\GSI}
\author{R.~Bassini}		\affiliation{\MIL}
\author{M.~Begemann-Blaich}	\affiliation{\GSI}
\author{A.~Ferrero}		\affiliation{\MIL}
\author{S.~Fritz}		\affiliation{\GSI}
\author{S.J.~Gaff-Ejakov}	\affiliation{\MSU}
\author{C.~Gro{\ss}}		\affiliation{\GSI}
\author{G.~Imm\'{e}}		\affiliation{\CAT}
\author{I.~Iori}		\affiliation{\MIL}
\author{U.~Kleinevo{\ss}}	\affiliation{\GSI}
\author{G.J.~Kunde}		\affiliation{\MSU}\affiliation{\LANL}
\author{W.D.~Kunze}		\affiliation{\GSI}
\author{A.~Le~F{\`e}vre}	\affiliation{\GSI}
\author{V.~Lindenstruth}	\affiliation{\GSI}\affiliation{\KIP}
\author{J.~{\L}ukasik}          \affiliation{\GSI}\affiliation{\IFJ}
\author{U.~Lynen}		\affiliation{\GSI}
\author{V.~Maddalena}		\affiliation{\CAT}     
\author{M.~Mahi}		\affiliation{\GSI}
\author{T.~M{\"o}hlenkamp}	\affiliation{\ROSS}
\author{A.~Moroni}		\affiliation{\MIL}
\author{W.F.J.~M{\"u}ller}	\affiliation{\GSI} 
\author{C.~Nociforo}		\affiliation{\CAT}     
\author{B.~Ocker}		\affiliation{\IKF}
\author{T.~Odeh}		\affiliation{\GSI}
\author{H.~Orth}		\affiliation{\GSI}
\author{F.~Petruzzelli}		\affiliation{\MIL}
\author{J.~Pochodzalla}		\affiliation{\GSI}\affiliation{\MAINZ}
\author{G.~Raciti}		\affiliation{\CAT}
\author{G.~Riccobene}		\affiliation{\CAT}     
\author{F.P.~Romano}		\affiliation{\CAT}     
\author{Th.~Rubehn}		\affiliation{\GSI}
\author{A.~Saija}		\affiliation{\CAT}   
\author{H.~Sann}\thanks{deceased}		\affiliation{\GSI}
\author{M.~Schnittker}		\affiliation{\GSI}
\author{A.~Sch{\"u}ttauf}	\affiliation{\GSI}\affiliation{\IKF}
\author{C.~Schwarz}		\affiliation{\GSI}
\author{W.~Seidel}		\affiliation{\ROSS}
\author{V.~Serfling}		\affiliation{\IKF}
\author{C.~Sfienti}		\affiliation{\GSI}
\author{A.~Trzci\'{n}ski}	\affiliation{\SINS}
\author{A.~Tucholski}		\affiliation{\SINS}
\author{G.~Verde}		\affiliation{\CAT}
\author{A.~W{\"o}rner}		\affiliation{\GSI}
\author{Hongfei~Xi}		\affiliation{\GSI}
\author{B.~Zwiegli\'{n}ski}	\affiliation{\SINS}
\collaboration{The ALADIN Collaboration}

\date{\today}


\begin{abstract}

Isotope temperatures from double ratios of hydrogen, helium, lithium, beryllium,
and carbon isotopic yields, and excited-state temperatures
from yield ratios of particle-unstable resonances in $^{4}$He, 
$^{5}$Li, and $^{8}$Be, were determined for spectator fragmentation, following 
collisions of $^{197}$Au with targets ranging from C to Au at incident
energies of 600 and 1000 MeV per nucleon.
A deviation of the isotopic from the excited-state temperatures is observed
which coincides with the transition from residue formation to 
multi-fragment production, suggesting a 
chemical freeze-out prior to thermal freeze-out in bulk disintegrations.

\end{abstract}


\maketitle


\section{Introduction}
\label{sec:intro}

In a disintegration scenario, the products drop out of equilibrium
roughly when their interaction rate falls below the expansion rate.
A thermal freeze-out is 
commonly defined as occurring when scattering processes 
cease to be effective in redistributing the particle momenta. From that 
time on, apart from sequential decays, 
only the long-range Coulomb forces continue to modify the 
otherwise frozen momenta of the disintegration products. Observables 
depending on the relative momenta, such as two-particle correlation 
functions, are sensitive to the source properties at the thermal 
freeze-out time \cite{ardouin,gong91}. The relative population of 
two-particle resonances, in an equilibrium situation, will therefore 
reflect the temperature at that stage \cite{poch87,morri94}.

Chemical freeze-out, on the other hand, is reached once the partitioning 
into particles and fragments is completed, and mutual scatterings are
no longer sufficiently energetic to significantly modify the channel 
composition by transfer, breakup or capture-type processes. 
The statistical multifragmentation 
models assume global equilibrium and thus coinciding thermal and 
chemical freeze-out times \cite{gross,bond95,botv06}.
The question to what extent this is found to be
realized in nuclear fragmentation reactions has been addressed by several 
groups. For example, by explicitly taking into account the effects of 
sequential decay, Huang et al. were able to demonstrate that in central
$^{197}$Au on $^{197}$Au collisions at 35 MeV per nucleon the
temperatures of the thermal and chemical freeze-out stages are
$\approx$ 4.3 MeV and identical to within 200 keV \cite{huang97}. 
A similar observation had earlier been made for $^{36}$Ar on $^{197}$Au 
at the same energy per nucleon \cite{tsang96}.  

In the central $^{197}$Au on $^{197}$Au reactions, however, when the bombarding 
energy is increased, temperatures derived from the population of 
two-particle resonances and from double ratios of isotopic yields start to 
diverge significantly \cite{serf98}. The excited-state temperatures 
saturate at values near $T$ = 5 MeV while the isotope temperatures 
increase to more than 10 MeV at 200 MeV per nucleon, the latter behavior 
being more in line with what is expected from the increasing system excitation.
Differences of similar magnitude have also been reported for $^{86}$Kr on $^{93}$Nb
reactions, studied at incident energies up to 120 MeV per nucleon \cite{xi98}.
Calculations showed that sequential decay should alter the two types of 
temperatures in a coherent fashion and, therefore, 
cannot account for the observed effect \cite{xi98a}.

The divergence of isotope and excited-state temperatures 
in central $^{197}$Au on $^{197}$Au collisions develops at bombarding energies 
at which collective radial flow becomes a significant part of the kinetic
energies of light particles and fragments 
\cite{hsi94,williams97,reis97,lefevre04}. 
This coincidence was interpreted as indicating a connection with the loss of 
global equilibrium resulting from the collective expansion. Radial 
expansion is accompanied by a local cooling which favors cluster
formation by coalescence \cite{reisdorf04}, 
as also demonstrated with nuclear molecular dynamics 
calculations \cite{bond95a}, a process terminated by the rapid rarefaction 
of the outstreaming material. In this non-equilibrium scenario, the higher 
temperatures of earlier reaction stages may thus manifest themselves not 
only in the higher kinetic energies of the fragments but also in 
partitions representing higher chemical temperatures.

It will be shown in this work that a saturation of excited-state 
temperatures, while the isotope temperatures rise to higher values, 
is also observed in spectator fragmentations in which flow effects 
are less prominent \cite{kunde95,gait00,odeh00,avde02,turbide04}. 
The divergence starts when 
the residue formation prevailing at lower excitation 
energies gives way to the multiple production of intermediate-mass fragments
as more energy is deposited in the spectator system. 
This suggests that it is the more general condition of a bulk 
breakup, independent of the magnitude of the collective motion accompanying 
it, which leads to a separation between the chemical and thermal 
freeze-outs. 

Apart from this main conclusion, the paper is also
meant to provide a summary of different temperature measurements 
performed for spectator decays, here specifically for collisions of 
$^{197}$Au nuclei at incident energies of 600 and 1000 MeV per nucleon. 
Some of the results are already published 
\cite{serf98,odeh00,poch95,xi97}. Questions addressed in the 
following, but not necessarily discussed exhaustively, include 
the $Z_{\rm bound}$-scaling of spectator decays, the validity of the 
chemical freeze-out picture, the reliability of sequential-decay corrections,
and the magnitude of the internal fragment excitation.

\section{Experiments and analysis}
\label{sec:exp}

The temperature measurements reported here were performed in two experiments
devoted to spectator decays following collisions of 
$^{197}$Au projectiles with a variety of targets ranging 
from C to Au at bombarding energies of 600 and 1000 MeV per nucleon.
In the first experiment, the ALADIN spectrometer was used
to detect and identify the products of the projectile-spectator
decay \cite{poch95,schuett96}. The masses of light fragments were 
determined by measuring their charge, magnetic-rigidity and velocity.
In the second experiment which concentrated on the $^{197}$Au on $^{197}$Au 
system at 1000 MeV per nucleon, the target-spectator decay was studied 
at backward angles with two multi-detector hodoscopes covering the 
angular range $122^{\circ} \leq \theta_{\rm lab} \leq 156^{\circ}$ and 
with three high-resolution telescopes placed at 
$\theta_{\rm lab} = 110^{\circ}, 130^{\circ}$, and 150$^{\circ}$.
These detectors were used to measure the yields and correlations of 
isotopically resolved light charged particles and fragments. 
The hodoscopes consisted of a total 
of 160 Si-CsI(Tl) telescopes in closely-packed geometry while the 
telescopes each consisted of
three Si detectors with thickness 50, 300, and 1000 $\mu$m and
of a 4-cm long CsI(Tl) scintillator with photodiode readout. 
Further details and results obtained in the second experiment 
may be found in Refs.~\cite{serf98,odeh00,xi97,fritz99}.

In both experiments, the variable $Z_{\rm bound}$, as measured for the projectile
spectator with the ALADIN time-of-flight wall, was used for sorting the data 
according to impact parameter.  
$Z_{\rm bound}$ is defined as the sum of the atomic numbers $Z_i$ of 
all projectile fragments with $Z_i \geq$ 2. It reflects the variation of the 
charge of the primary spectator system and is monotonically
correlated with the impact parameter of the reaction \cite{ogilvie93}. 
In the second experiment, when the decay of the target spectator was studied,
the impact-parameter selection with the  $Z_{\rm bound}$ of the projectile
took advantage of the symmetry of the $^{197}$Au on $^{197}$Au system \cite{xi97}. 
 
The excited-state temperatures are deduced from the populations 
of particle-unstable resonances in emitted $^4$He, $^5$Li and $^8$Be  
nuclei, determined from two-particle coincidences measured with the 
Si-CsI hodoscopes.
The peak structures were identified by using the technique of correlation
functions, and background corrections were based on results obtained for
resonance-free pairs of light fragments.
The pair of the ground state (g.s.) and 16.66-MeV excited state of $^5$Li,
identified in p-$^4$He and d-$^3$He coincidence yields,
represents a widely used thermometer for nuclear
reactions \cite{poch87,kunde91,schwarz93}. In addition,
correlated yields of p-t, p-$^7$Li, and $^4$He-$^4$He coincidences and
$^4$He single yields were used to deduce temperatures from the
relative populations of states or groups of states
in $^4$He (g.s.; 21.21 MeV and higher)
and in $^8$Be (3.04 MeV; 17.64 and 18.15 MeV).
The relative efficiencies for the correlated detection
of the decay products were calculated with a
Monte Carlo model \cite{kunde91,serf97,fritz97}. The estimated
uncertainties of the background functions are the main
contributions to the errors of these excited-state temperatures.

The isotope temperatures were deduced from the double ratios of yields of 
neighboring isotopes, as described previously \cite{poch95,xi97,isobook}. 
Energy-integrated yields were measured for the projectile decay for which
the acceptance of the ALADIN spectrometer provides good solid-angle 
coverage. For the target spectator, isotopically resolved yields 
were obtained from the four-element telescopes.
Besides the commonly used $T_{\rm HeLi}$ thermometer, based on the 
$^3$He/$^4$He and $^6$Li/$^7$Li yield ratios, temperature observables deduced 
from other pairs of isotopes were also evaluated. 
The corresponding expressions are

\begin{equation}
T_{\rm HeLi} = 13.3 MeV/\ln(2.2\frac{Y_{^{6}{\rm Li}}/Y_{^{7}{\rm Li}}}
{Y_{^{3}{\rm He}}/Y_{^{4}{\rm He}}}).
\label{eq:heli}
\end{equation}
 
\begin{equation}
T_{\rm Hepd} = 18.4 MeV/\ln(5.5\frac{Y_{\rm p}/Y_{\rm d}}
{Y_{^{3}{\rm He}}/Y_{^{4}{\rm He}}}).
\label{eq:hepd}
\end{equation}

\begin{equation}
T_{\rm Hedt} = 14.3 MeV/\ln(1.6\frac{Y_{\rm d}/Y_{\rm t}}
{Y_{^{3}{\rm He}}/Y_{^{4}{\rm He}}}).
\label{eq:hedt}
\end{equation}

\begin{equation}
T_{\rm BeLi} = 11.3 MeV/\ln(1.8\frac{Y_{^{9}{\rm Be}}/Y_{^{8}{\rm Li}}}
{Y_{^{7}{\rm Be}}/Y_{^{6}{\rm Li}}}).
\label{eq:beli}
\end{equation}

\begin{equation}
T_{\rm tHeLiBe} = 14.2 MeV/\ln(2.2\frac{Y_{^{6}{\rm Li}}/Y_{^{7}{\rm Be}}}
{Y_{\rm t}/Y_{^{4}{\rm He}}}).
\label{eq:thelibe}
\end{equation}

\begin{equation}
T_{\rm CLi} = 11.5 MeV/\ln(5.9\frac{Y_{^{6}{\rm Li}}/Y_{^{7}{\rm Li}}}
{Y_{^{11}{\rm C}}/Y_{^{12}{\rm C}}}).
\label{eq:cli}
\end{equation}

\begin{equation}
T_{\rm CC} = 13.8 MeV/\ln(7.9\frac{Y_{^{12}{\rm C}}/Y_{^{13}{\rm C}}}
{Y_{^{11}{\rm C}}/Y_{^{12}{\rm C}}}).
\label{eq:cc}
\end{equation}

The prefactors in Eqs.~(\ref{eq:heli})-(\ref{eq:cc}) represent
the double differences of the binding energies of the four isotopes
appearing in the double yield ratios. They are between 11 and 19 MeV 
and thus fulfill the requirement that this quantity should
be large compared to the anticipated temperatures \cite{albergo,tsang97}.
The numerical factors preceding the double yield ratios in the denominator
contain the mass numbers and the ground-state degeneracies 
2$s_{\rm g.s.}$+1 of these isotopes \cite{albergo}.
The expressions are thus strictly valid only for the ground-state populations
in thermal equilibrium. Since they may be subsequently altered by the 
feeding from excited states and from particle-unbound resonances, 
temperatures obtained with these expressions from measured yields are
referred to as apparent temperatures.
The magnitude of the corrections required to account for these effects was estimated
on the basis of calculations with the Quantum Statistical Model 
of Hahn and St{\"o}cker \cite{hahn88} in the version of Konopka et al. 
\cite{konop94}, but also with other statistical models
as described previously \cite{poch95,xi97} and discussed below.

\begin{figure}		
     \epsfysize=12.0cm
     \centerline{\epsffile{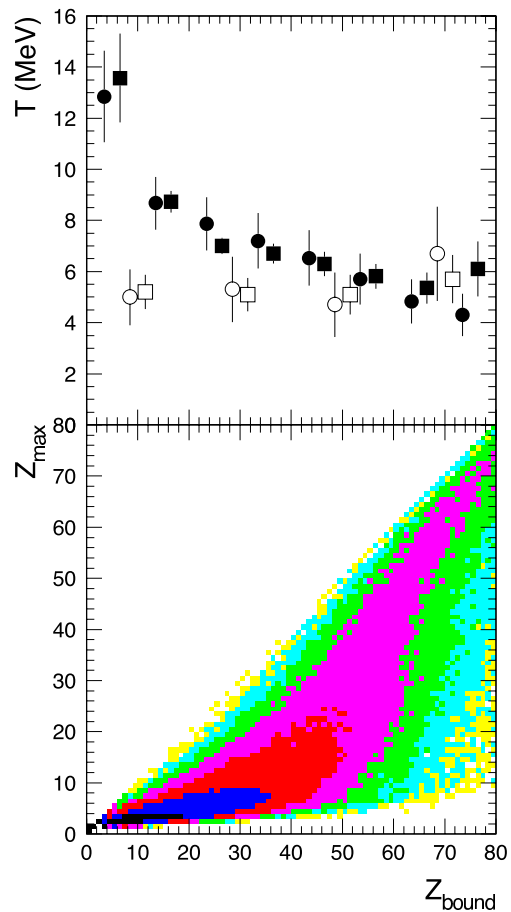}}
\caption{\label{fig:zbound} (Color online)
Results obtained for $^{197}$Au on $^{197}$Au at $E/A$ = 1000 MeV:
Top: Measured isotope temperature $T_{\rm HeLi}$ (full circles and
squares representing the data for the target and projectile decays, 
respectively)
and excited-state temperatures (open circles and squares representing the 
data for $^5$Li and $^4$He, respectively) as a function of 
$Z_{\rm bound}$. Average values for 10 or 20 unit intervals of $Z_{\rm bound}$
are given. For clarity, the data symbols are slightly displaced horizontally.
The indicated uncertainties are mainly of systematic origin.
Bottom: Distribution of $Z_{\rm max}$ versus $Z_{\rm bound}$ after
conventional fission events have been removed. The
shadings follow a logarithmic scale.
}
\end{figure}

\section{Results}
\label{sec:resu}

Temperature values obtained for 
$^{197}$Au on $^{197}$Au at 1000 MeV per nucleon are summarized in 
Fig. \ref{fig:zbound}, top, 
where they are displayed as a function of $Z_{\rm bound}$.
The closed symbols represent the isotope temperature $T_{{\rm HeLi}}$ 
as derived for the target spectator from the energy-integrated yields
measured at $\theta_{\rm lab} = 150^{\circ}$ 
and for the projectile spectator from the measured total yields.
They include a correction factor 1.2 to account for the effects
of secondary decays which was obtained as an average over
the predictions of several statistical models \cite{poch95,xi97}.
The two data sets agree within errors as expected because of the symmetry 
of the reaction. The different solid-angle coverage and detection techniques, 
apparently, do not cause measurable differences.

The open symbols represent the temperatures deduced from the ratios of 
state populations in $^4$He and $^5$Li, integrated over 20-unit intervals 
of $Z_{\rm bound}$ for reasons of counting statistics. 
In striking contrast to the isotope temperature $T_{{\rm HeLi}}$ 
which monotonically rises up to about 12 MeV,
they are virtually independent of $Z_{\rm bound}$, mutually
consistent with each other, and
define a mean value of $\approx$ 5 MeV. 
A temperature $T_{{\rm ^8Be}}$ = 5.6 $\pm$ 1.2 MeV, 
averaged over the full range of $Z_{\rm bound}$ because of low count rates,
was derived from the ratio of excited-state populations in $^8$Be.
A similar behavior was observed for central $^{197}$Au on $^{197}$Au collisions
when the energy was raised from 50 up to 200 MeV per nucleon~\cite{serf98}.

An important characteristics of the spectator decay
is illustrated in Fig. \ref{fig:zbound}, bottom. 
It shows the event distribution in the
$Z_{\rm max}$-versus-$Z_{\rm bound}$ plane, where $Z_{\rm max}$ denotes the largest
atomic number $Z$ observed in the event. At large $Z_{\rm bound}$,
corresponding to small excitation energies, $Z_{\rm max}$ is very close to
$Z_{\rm bound}$, indicating that mainly one large fragment or heavy residue
is produced. 
At small $Z_{\rm bound}$ and high excitations, 
$Z_{\rm max}$ is much smaller than $Z_{\rm bound}$ which implies
that the disintegration has additionally produced several other fragments, 
as equally evident from the measured multiplicity 
distributions \cite{schuett96}.
The transition between these two regimes, at $Z_{\rm bound}$ between 50 and 60,
is marked by a fairly rapid
drop of the ridge line representing the most probable $Z_{\rm max}$. 

Using a simplified statistical model for multifragmentation, 
Das Gupta and Mekjian have shown that the decrease of the relative mass 
of the largest fragment with respect to the system mass, $A_{\rm max}/A_0$, 
is rather sharp if regarded on the temperature scale \cite{dasg98}. 
At this temperature, termed "boiling temperature" by the authors, a peak in 
the specific heat develops which becomes increasingly pronounced 
as the system mass $A_0$ is increased to large values, $A_0 > 1000$, i.e., 
far beyond the limits of nuclear masses. 
For $A_0$ = 150, the system mass corresponding to 
$Z_{\rm bound}\approx$ 50~\cite{poch95}, 
the predicted specific heat distribution is wider with a maximum at $T = 6.3$~MeV.
This is slightly larger than the value $T_{\rm HeLi} = 5.8 \pm 0.8$ MeV
measured for $50 < Z_{\rm bound} < 60$
but, within errors, consistent with it. In this respect, a link may be 
seen between the observed onset of multi-fragment disintegrations and the 
first-order phase transition expected for large nuclear systems. The
two temperature observables, incidentally, start to deviate from 
each other in this transition region.

\begin{figure}		
     \epsfysize=12.0cm
     \centerline{\epsffile{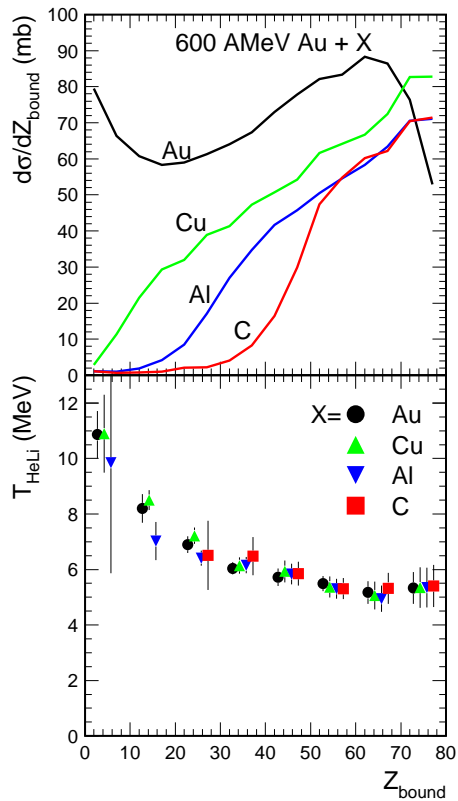}}
\caption{\label{fig:target} (Color online)
Top:
Measured cross sections $d\sigma /dZ_{\rm bound}$
for the reactions of $^{197}$Au projectiles 
at $E/A$ = 600 MeV with targets
of C, Al, Cu, and Au.
Note that the experimental trigger
affected the cross sections for $Z_{\rm bound} \geq$ 65. 
Bottom:
Isotope temperature $T_{\rm HeLi}$ measured for the same four reactions
presented as averages over 10-unit wide bins of $Z_{\rm bound}$.
For clarity, the data symbols are slightly displaced horizontally.
}
\end{figure}

The invariance of $Z_{\rm bound}$-sorted fragment multiplicities and related
observables describing the partitioning of the system represents
a prominent characteristics of spectator fragmentation. 
The validity of $Z_{\rm bound}$ scaling with respect to the choice of the target
was one of the early indications for equilibrium being reached 
at breakup \cite{hubele91,kreutz93}.
This invariance property is also shared by the chemical freeze-out 
temperature $T_{\rm HeLi}$ as shown in Fig. \ref{fig:target}, bottom,
for the projectile decay at 600 MeV per nucleon after collisions with
target nuclei ranging from carbon to gold. 
The mass of the target only determines which part of the $Z_{\rm bound}$ range 
will be populated in the reaction (Fig.~\ref{fig:target}, top). Heavy targets 
are required for initiating the most violent collisions associated with
small $Z_{\rm bound}$.

\begin{figure}		
     \epsfxsize=0.9\columnwidth
     \centerline{\epsffile{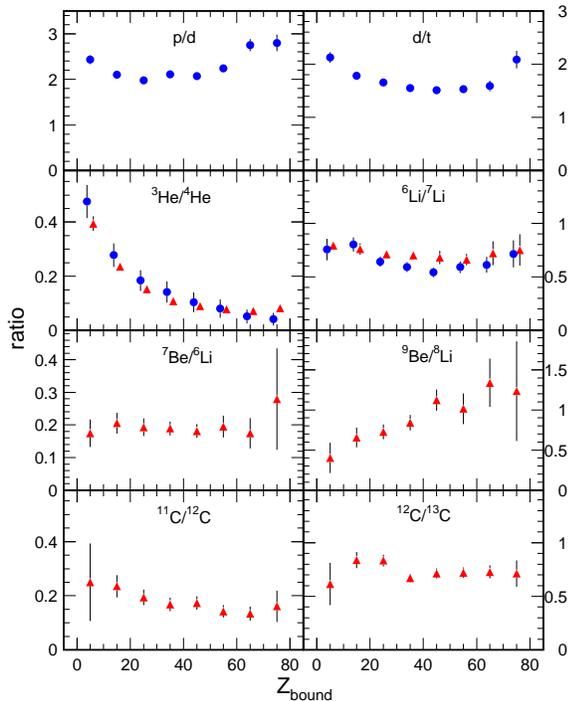}}
\caption{\label{fig:ratios} (Color online)
Measured isotope yield ratios for $^{197}$Au on $^{197}$Au reactions 
as a function of $Z_{\rm bound}$. Results obtained for projectile fragments
at 600 MeV per nucleon and for target fragments at 1000 MeV per nucleon 
incident energy are represented by triangles and circles, respectively.
}
\end{figure}

The thermometry with the $^{3,4}$He isotope pairs benefits from the large 
difference $\Delta B =$ 20.6 MeV of the binding energies of these two nuclei 
but it is not the only choice. 
There are also other combinations of isotopes which can be expected to
provide the necessary sensitivity for the measurement of temperatures 
in the MeV range. The additional cases studied here are the
thermometers $T_{\rm BeLi}$, $T_{\rm tHeLiBe}$, $T_{\rm CLi}$, and $T_{\rm CC}$ 
(Eqs.~(\ref{eq:beli})-(\ref{eq:cc})). In $T_{\rm tHeLiBe}$, $^3$He is replaced
by the triton which has a very similar binding energy, and consequently 
also $^7$Li by $^7$Be. The $T_{\rm BeLi}$ thermometer is interesting because it
is exclusively based on fragments with $Z \geq 3$ and not on 
light-particle yields. With the carbon thermometers $T_{\rm CLi}$ and $T_{\rm CC}$
one hopes to exploit the large difference of the  $^{11,12}$C binding energies of 
$\Delta B =$ 18.7 MeV. Their use can permit a test of chemical equilibrium up 
to this range of fragment masses.
 
Measured isotope ratios, as needed for these thermometers, are
shown in Fig.~\ref{fig:ratios}. For the $^{3,4}$He and $^{6,7}$Li isotope pairs,
ratios obtained for the integrated yields of projectile fragments at 600
MeV per nucleon and for the yields of target fragments 
measured at $\theta_{\rm lab} = 150^{\circ}$ and 1000 MeV 
per nucleon agree within errors.
Since this is again a manifestation of $Z_{\rm bound}$ scaling, 
here of the invariance with bombarding energy \cite{schuett96,xi97}, 
the distinction between these cases will be dropped in the following. 
Of the eight yield ratios shown, only the $^{3,4}$He
ratio and, less pronounced, also the $^{9}$Be/$^{8}$Li ratio 
($\Delta B =$ 16.9 MeV) vary strongly with $Z_{\rm bound}$. 
For $^{11,12}$C, the total
variation amounts to hardly a factor of 2, less than the factor 3 for
$^{9}$Be/$^{8}$Li, while the other ratios are rather flat. For these remaining 
five pairs of isotopes, this is not unexpected because the differences of 
their binding energies are within $\Delta B =$ 2.2 MeV (p,d) and $\Delta B =$ 7.3 MeV
($^{6,7}$Li).

\begin{figure}		
     \epsfysize=7.0cm
     \centerline{\epsffile{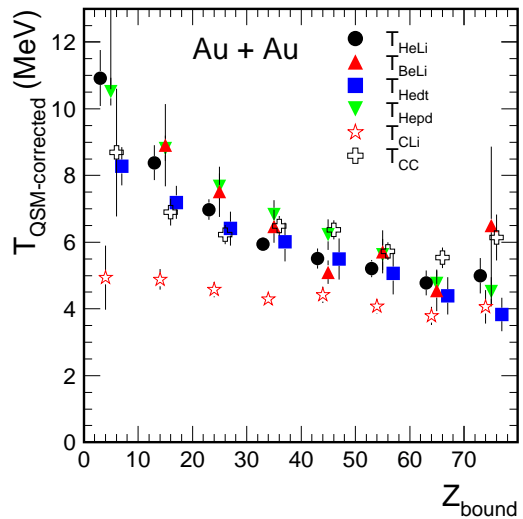}}
\caption{\label{fig:qsm} (Color online)
Isotope temperatures for $^{197}$Au on $^{197}$Au at 600 and 1000 MeV per nucleon,
with corrections for the effects of secondary decay according to the
quantum statistical model \protect\cite{hahn88,konop94},
as a function of $Z_{\rm bound}$.
}
\end{figure}

Results for six temperatures obtained with these ratios are shown in Fig.~\ref{fig:qsm}. 
They have been corrected for the effect of
secondary decays by using the predictions of the quantum statistical
model~\cite{hahn88,konop94},
calculated for a breakup density $\rho /\rho_0$ = 0.3. 
With the exception of $T_{\rm CLi}$, all temperatures increase slowly
as the collisions pass through the rise and fall 
of fragment production with decreasing $Z_{\rm bound}$. 
The values are about 4 to 6 MeV for peripheral collisions and reach 8 to 10 MeV for 
the most violent fragmentations at small impact parameters. 
The rise is, notably, also observed with $T_{\rm BeLi}$ and 
thus not a singular consequence of the behavior of the $^{3,4}$He yield ratio.
Only the carbon-lithium thermometer does not fit into this otherwise rather 
consistent picture. Here the small variations of the $^{11,12}$C and $^{6,7}$Li
yield ratios, apparently, compensate each other (Fig.~\ref{fig:ratios}), leading to
a nearly constant temperature which is not changed much by the corrections. At very 
large and very small $Z_{\rm bound}$, at the boundaries of the fragmentation regime,
the errors are large for temperatures deduced from beryllium and carbon isotopes
because the yields are small. 
 
\section{Discussion}
\label{sec:disc}

\subsection{Target invariance}
\label{sec:inv}

Reports of a saturation of the breakup temperature
at high excitation energies in $^{197}$Au on $^{12}$C at 1000 MeV per 
nucleon \cite{hauger98} and of a difference between this 
and the $^{197}$Au on $^{197}$Au reactions have led to discussions in the 
literature \cite{majka97} and have inspired theoretical investigations into their 
possible origin \cite{samad97}.
In the present data, no significant difference is observed 
for reactions of the $^{197}$Au projectiles with different targets if 
the temperature is viewed as a function of $Z_{\rm bound}$ (Fig. \ref{fig:target}).
As the $Z_{\rm bound}$ region populated with the lighter targets is 
limited, comparisons will have to be restricted to the common range.
In the data of the EOS collaboration \cite{hauger98,schar01}, with 
$^{197}$Au on $^{12}$C covering the rise of fragment production,
the temperature $T_{\rm Hedt}$, e.g., rises
from about 4 to 6 MeV with increasing centrality. This is very similar
to the results obtained here for $Z_{\rm bound} > 40$, the corresponding interval 
in $^{197}$Au on $^{197}$Au (Fig.~\ref{fig:qsm}). The corrections are small
for $T_{\rm Hedt}$ (see below) which permits a direct comparison 
of the two data sets.

\begin{figure}		
     \epsfysize=9.0cm
     \centerline{\epsffile{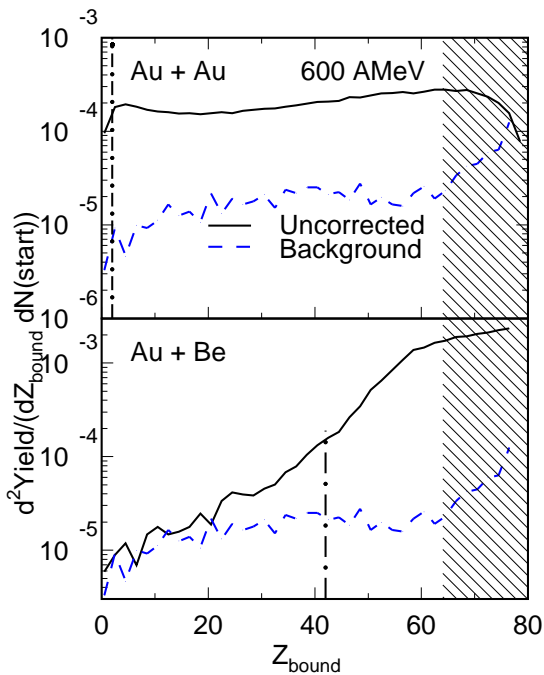}}
\caption{\label{fig:sigma}
Cross sections $d\sigma /dZ_{\rm bound}$ (in units of events per projectile)
for the reactions of $^{197}$Au projectiles
at $E/A$ = 600 MeV with the two targets
of Au (top) and Be (bottom) before background subtraction (full line)
and background generated by upstream detectors (dashed). The dash-dotted 
lines indicate the value of $Z_{\rm bound}$ at which the cross-section integral 
has reached a relative value of 1.5\%. Inefficiencies of the experimental 
trigger affect the range $Z_{\rm bound} \geq$ 65 (hatched).
}
\end{figure}

In the present studies of projectile fragmentation, the  
apparent tails of the cross section d$\sigma$/d$Z_{\rm bound}$ toward smaller 
values of $Z_{\rm bound}$ (Fig. \ref{fig:target}, top) can be 
quantitatively assigned to background reactions in thin plastic detectors 
mounted about 1.5 m upstream of the target. These detectors containing
scintillator material with a total areal density of typically 15 mg/cm$^2$ 
were used to measure the arrival time and transverse coordinates of the 
incoming beam particles. The ALADIN spectrometer has an increasingly reduced
acceptance for reactions occurring further upstream. In reactions occurring there,
some of the produced projectile fragments may, therefore, escape detection, 
and the recorded $Z_{\rm bound}$ will be too small. 
The yield of reactions in these foils, measured by removing
the regular target, is shown in Fig. \ref{fig:sigma} in 
comparison with the cross sections d$\sigma$/d$Z_{\rm bound}$ measured for reactions of 
the $^{197}$Au projectiles with the regular targets of $^{197}$Au 
(480 mg/cm$^2$) and $^9$Be (700 mg/cm$^2$). 
The figure demonstrates that the population of the region 
$Z_{\rm bound} <$ 30, corresponding to excitations of $E/A >$ 10 MeV \cite{poch95},
is negligible if a beryllium target is used. The tail of the cross section
at $Z_{\rm bound} <$ 45 ($E/A >$ 8 MeV) is on the percent level and may even 
be partly due to fluctuations caused by other small experimental
inefficiencies and 
by secondary reactions within the target and detectors.
In the analysis of the present experiments, the first 1.5\% of the 
integrated reaction cross section at small $Z_{\rm bound}$ in the 
corrected spectrum (to the left of the dashed-dotted lines in 
Fig. \ref{fig:sigma}) have been ignored (see also Ref.~\cite{schuett96}).

\subsection{Corrections}
\label{sec:corr}

The need to consider the effects of secondary decay is known since 
breakup temperatures are being measured and was respected in the
study of $^{197}$Au fragmentations \cite{poch95}. 
Besides the ground states considered in 
Eqs.~(\ref{eq:heli})-(\ref{eq:cc}) also excited states are expected to be
populated 
according to the equilibrium conditions and both may be fed 
by particle decays of heavier products. The problem has been addressed 
by several groups and, besides the work already cited 
(Refs.~\cite{poch87,huang97,xi98,tsang96,xi98a,xi97,tsang97,majka97}),
the reader is referred to Refs.~\cite{chen88,gulm97,xi99,viola99,kelic06}
and references cited therein.

\begin{figure}		
     \epsfxsize=\columnwidth
     \centerline{\epsffile{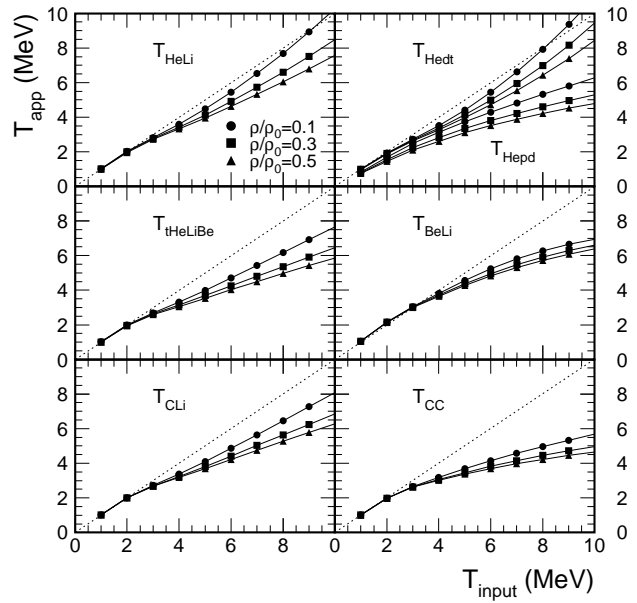}} 
\caption{\label{fig:qsmcorr}
Apparent temperatures $T_{\rm app}$ as a function of the temperatures 
$T_{\rm input}$ used in calculations with the quantum-statistical model of 
Refs.~\protect\cite{hahn88,konop94} (full lines) for the seven temperature 
observables of Eqs.~(\protect\ref{eq:heli})-(\protect\ref{eq:cc}).
The symbols distinguish
three assumptions for the breakup density as indicated in the left top panel.
The dotted lines mark the diagonal $T_{\rm app} = T_{\rm input}$.
}
\end{figure}

In the following, the corrections for the isotope temperatures 
predicted by the quantum-statistical model of Hahn and 
St{\"o}cker \cite{hahn88,konop94} will be discussed in more detail.
This model assumes thermal and chemical equilibrium at breakup at which
the fragmenting system is characterized by a density $\rho$, temperature $T$, 
and by its overall $N/Z$ ratio. Fermion and boson statistics are respected
which, however, is not crucial at high temperature. 
The model does not consider the finite size of nuclear systems but follows 
the sequential decay of excited fragments according to tabulated 
branching ratios. It is exact in this respect, 
within the limits of its associated mass table.

Results obtained with this model for the seven temperature 
observables given in Eqs.~(\ref{eq:heli})-(\ref{eq:cc}) are shown
in Fig.~\ref{fig:qsmcorr}. The apparent temperatures $T_{\rm app}$, as
deduced from the formulas, are shown as a function of the equilibrium
temperatures $T_{\rm input}$ of the fragmenting system for which 
breakup densities between $\rho=0.1~\rho_0$ and $\rho=0.5~\rho_0$ were assumed.
At low temperatures, the
population of higher states is less important, and 
$T_{\rm app}$ is very close to $T_{\rm input}$ as expected. The deviations at
higher temperatures $T > 3$~MeV are smallest for the two observables
$T_{\rm HeLi}$ and $T_{\rm Hedt}$ but not negligible. For $T_{\rm HeLi}$ at 
$T_{\rm input} = 6$~MeV, the apparent $T_{\rm HeLi}$ is between 4.5 and 5.5 MeV, 
depending on the density. The deviations are generally larger for the 
higher densities. For $T_{\rm HeLi}$ and 
$\rho=0.3~\rho_0$, the predicted correction is well described
by $T_{\rm input} = 1.2 \cdot T_{\rm app}$, 
suggesting a global correction factor 1.2. 
A similar value has been derived as an average from 
calculations with several statistical 
models \cite{poch95}, including GEMINI~\cite{charity88} and the Berlin
microcanonical multifragmentation model MMMC~\cite{gross}. It was used 
for $T_{\rm HeLi}$ in previous work 
(Refs.~\cite{serf98,odeh00,poch95,xi97,lefevre05}) and applied also here to 
obtain the results shown in Figs.~\ref{fig:zbound} and \ref{fig:target}. 
A correction of 20\% is indicated also by calculations with the Copenhagen
Statistical Fragmentation Model as shown in Ref.~\cite{xi97}. 

The predicted deviations of $T_{\rm app}$ from $T_{\rm input}$ 
at higher temperatures are 
considerably larger for the other observables, notably for 
$T_{\rm Hepd}$ and $T_{\rm CC}$; small differences
of $T_{\rm app}$ correspond to large differences of $T_{\rm input}$ which makes
these temperatures less precise. This is reflected by the 
errors adopted for the results shown in Fig.~\ref{fig:qsm} 
for which the QSM predictions for $\rho=0.3~\rho_0$ 
were used for the corrections.
Calculations were also performed with the model of 
Tsang, Xi et al. \cite{tsang97,xi99} which is partly adapted to binary emissions
and has been found to consistently 
reproduce observed variations of $T_{\rm app}$ in several reactions, 
including the $^{197}$Au on $^{197}$Au reaction 
at 35 MeV per nucleon cited above \cite{huang97}. 
The results are qualitatively similar to those obtained with the QSM except
for $T_{\rm CLi}$ and $T_{\rm CC}$ for which much smaller corrections are obtained
(for details see \cite{odeh99}). In these cases, 
the predicted apparent temperatures seem to depend very sensitively 
on the model assumptions.

\begin{figure}		
     \epsfxsize=\columnwidth
     \centerline{\epsffile{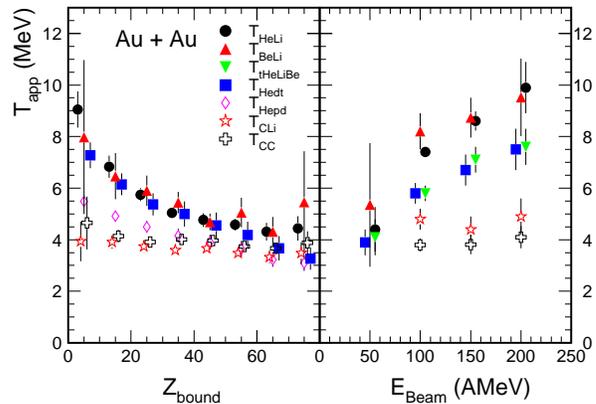}}
\caption{\label{fig:alltemp}  (Color online)
Apparent isotope temperatures 
(Eqs.~(\protect\ref{eq:heli})-(\protect\ref{eq:cc})) 
in $^{197}$Au on $^{197}$Au reactions for spectator decays 
at 600 and 1000 MeV per nucleon as a function of $Z_{\rm bound}$ 
(left panel) and for central collisions at 50 to 200 MeV per 
nucleon as a function of the incident energy (right panel).
}
\end{figure}

The temperature observables associated with the largest predicted
uncertainties, $T_{\rm Hepd}$, $T_{\rm CLi}$ and $T_{\rm CC}$, are incidentally 
also those with the largest deviations from the common behavior of 
the others. Their apparent values depend very little on $Z_{\rm bound}$ and 
those involving carbon isotope ratios essentially saturate, 
an observation also made for the central $^{197}$Au on $^{197}$Au collisions 
at 50 to 200 MeV per nucleon (Fig.~\ref{fig:alltemp}, right panel, Ref.~\cite{serf98}).
According to the QSM, this behavior is 
expected for $T_{\rm Hepd}$ and $T_{\rm CC}$, and the calculated corrections
suffice to restore the consistency with the others. This is not the case for
$T_{\rm CLi}$ for which only small corrections are predicted.
According to the model of Tsang, Xi et al., the corrections should be small
for both temperatures based on the $^{11,12}$C yield ratio, leading to
similar carbon temperatures of approximately 4 MeV, with no significant 
$Z_{\rm bound}$ dependence.
In either case, the concept of a common freeze-out temperature is not supported 
by the isotope temperatures deduced from carbon isotope yields.

The three excited-state temperatures are characterized by large energy 
differences of the considered states, and, as done previously, 
no corrections for sequential 
feeding were applied \cite{serf98,xi98}. This is best 
justified for the case of the particle-unstable resonances of $^5$Li 
\cite{poch87,fritz97,chen88} and found valid also for $^4$He as long as the 
temperature is low ($\approx 4$~MeV or less, cf. \cite{huang97}). However,  
side-feeding contributions to the measured $\alpha$ particle yields should 
have similar consequences for both, the excited-state and double-isotope 
temperatures, as discussed in Ref.~\cite{xi98a}. Applying the correction
factor of 1.2 used for $T_{\rm HeLi}$ will raise the $^4$He 
temperature to about 6~MeV but will not modify its invariance with respect to
$Z_{\rm bound}$ (Fig.~\ref{fig:zbound}). Within errors, also the consistency 
with the $^5$Li temperature will be maintained. The $^8$Be thermometer
requires a side-feeding correction of 1.1 $\pm$ 0.1, according to the 
calculations~\cite{fritz97},
which will raise the $Z_{\rm bound}$-averaged result to 6.2 $\pm$ 1.5 MeV. 

The observed saturation of the excited-state temperatures has interesting 
consequences also for the side-feeding corrections of the isotope temperatures. 
Generally, one would expect that the corrections are less important 
if the emerging fragments are less excited. The corrections for higher 
temperatures may, therefore, be overestimated if chemical and thermal equilibrium 
at breakup is assumed as in the QSM. However, if the chemical freeze-out precedes 
the thermal freeze-out, as the divergence of the temperatures suggests, 
the evolution of the system from one to the other will still have to be 
accounted for. There, the preformed fragments are unlikely to have reached 
their asymptotic quantum 
structure, and standard decay probabilities and branching ratios would not 
necessarily be adequate for that purpose. In-medium properties of fragments
in the hot environment are, in fact, attracting considerable interest presently,
with modifications being indicated by isotopic effects observed in recent studies
of fragment production (see, e.g., 
Refs.~\cite{lefevre05,ono04,botv06a,colonna06,shetty07} 
and references therein).

The order of magnitude of the effects to be expected may perhaps be illustrated 
by applying correction factors obtained for $T \approx 5$~MeV, the saturation 
temperature for internal excitations, to chemical temperatures higher than that 
value. More precisely, in the terminology introduced in 
Ref.~\cite{tsang97}, the calculated correction factor $\kappa$ for the double 
yield ratio at $T_{\rm app} = 5$~MeV is used for correcting 
all $T_{\rm app} \ge 5$~MeV.
With this procedure, the values for, e.g., $T_{\rm BeLi}$ at
$Z_{\rm bound} \approx 15$~and 25 (Fig.~\ref{fig:qsm})
will be reduced by 1.2 MeV and 0.6 MeV, respectively. 
The values for $T_{\rm HeLi}$ will increase, as shown in Table~\ref{tab:lnkb} 
where they are compared to the other correction methods discussed up to now. 
For completeness, also the temperatures obtained with the empirical factor $\kappa$ 
deduced for central $^{197}$Au on $^{197}$Au collisions at 35 MeV per nucleon 
($T \approx 4.3$~MeV, Ref.~\cite{xi98a}) are given in the table (last column). 
These values are even slightly smaller than $T_{\rm app}$ by up to 5\% and 
should probably be taken as lower limits (the corrected temperature should be 
larger than $T_{\rm app}$ if the sequential production of $^4$He is strong). 
Altogether, these results may be considered as representative for 
the increasing systematic uncertainty as the temperature to be measured increases.

\begin{table}
\caption{\label{tab:lnkb}
Corrected double-isotope temperatures $T_{\rm HeLi}$ as
obtained with four methods, (a) using the QSM predictions for $\rho=0.1~\rho_0$  
as shown in Fig.~\protect\ref{fig:qsmcorr}, (b) using the global correction 
factor 1.2 which is representative for the QSM results with $\rho=0.3~\rho_0$, 
(c) by applying the QSM correction factor for 5~MeV ($0.1~\rho_0$) 
for $T_{\rm app} \ge 5$~MeV, 
and (d) by using the empirical correction factor for 4.3 MeV 
of Ref.~\protect\cite{xi98a}. All values are given in MeV.
}
\begin{ruledtabular}
\begin{tabular}{l c c c c}
$T_{\rm app}$ & $T_{\rm QSM(0.1\rho_0)}$ & $1.2 \cdot T_{\rm app}$ & $T_{\rm QSM(5~MeV)}$ & $T_{\rm MSU(4.3~MeV)}$ \\
\hline
~5 & 5.6 & ~6.0 & ~5.7 & 4.9 \\
~6 & 6.6 & ~7.2 & ~7.0 & 5.8 \\
~7 & 7.4 & ~8.4 & ~8.5 & 6.8 \\
~8 & 8.2 & ~9.6 & 10.0 & 7.7 \\
~9 & 9.1 & 10.8 & 11.6 & 8.6 \\
\end{tabular}
\end{ruledtabular}
\end{table}

\subsection{Kinetic energies}
\label{sec:ekin}

\begin{figure}		
     \epsfxsize=\columnwidth
     \centerline{\epsffile{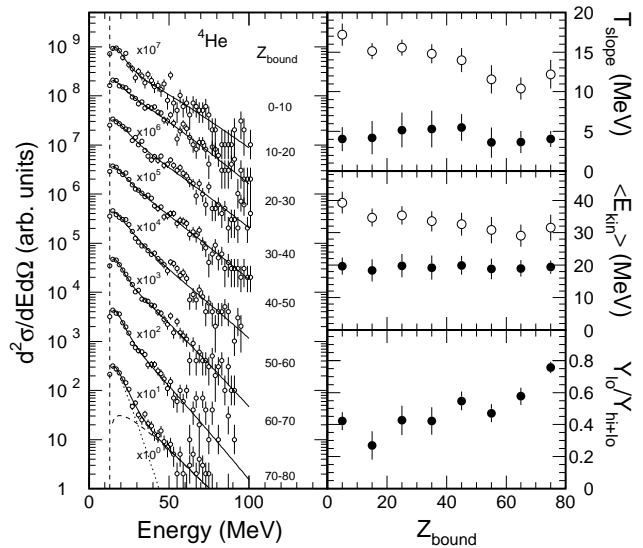}}
\caption{\label{fig:alphas}
Energy spectra of $^4$He measured at $\theta_{\rm lab} = 135^{\circ}$ for the reaction $^{197}$Au on $^{197}$Au at 1000 MeV per nucleon 
for eight intervals of $Z_{\rm bound}$  
(left panel). The full lines represent the 
results of a fit with two Maxwellian sources whose individual contributions 
are shown for the bin of largest $Z_{\rm bound}$ (dotted and dashed 
lines). The vertical dashed line indicates the threshold for identification.
The obtained slope temperatures, mean kinetic energies and relative strengths
of the high and low-temperature components are represented by the open and
closed circles, respectively (right panels). 
}
\end{figure}

In the kinetic-energy spectra of light products 
from the $^{197}$Au on $^{197}$Au reaction, a soft component 
characterized by temperatures of the order of 5 MeV or lower is 
well recognized for protons, neutrons, and $\alpha$ particles. 
For protons and neutrons, this has been shown in~\cite{odeh00,zwieg01,zwieg02}. 
For the case of $\alpha$ particles from the target-spectator decay 
at 1000 MeV per nucleon, this is illustrated in Fig.~\ref{fig:alphas}. 
Besides the measured spectra, the figure shows the slope temperatures, 
mean kinetic energies and relative intensities obtained from a 
two-component fit with Maxwellian sources. The low-temperature component
dominates for large $Z_{\rm bound}$ but drops to a strength of about 40\% for
smaller $Z_{\rm bound}$. Interpreting it as resulting from secondary 
emissions is consistent with the evolution of the reaction process
(Fig.~\ref{fig:zbound}). Below the fragmentation threshold 
(large $Z_{\rm bound}$), the observed $\alpha$ particle yields are expected to be
mainly produced by evaporation from excited heavy residues.  
In the domain of multifragmentation, secondary emission from heavier fragments 
should contribute only a fraction of the total yield. 
The fitted slope temperatures 
of 4 to 5~MeV are compatible with either process, evaporation as well as
secondary emission from nuclei in thermal equilibrium at the break-up 
temperature.

With the assumption that the identified low-temperature component is the 
main reason for the need of corrections, He-Li isotope temperatures can be 
derived using only the high-temperature yields of $\alpha$ particles without
further corrections. The results for this new temperature $T_{\rm hi}$
are given in Table~\ref{tab:thi}. The effective correction 
factors $R_{\rm hi}= T_{\rm hi}/T_{\rm app}$ with respect to the apparent temperature,
as obtained from the energy-integrated yields according to Eq.~(\ref{eq:heli}),
are between 1.3 and 1.4 and higher than the corrections  
for $T_{\rm HeLi}$ used up to now (Table~\ref{tab:lnkb}). 
Accordingly, the resulting $T_{\rm hi}$ is about 1~MeV higher
than the temperatures given in 
Figs.~\ref{fig:zbound},\ref{fig:target},and~\ref{fig:qsm}. 
In the bin of largest $Z_{\rm bound}$, the
relative correction is even larger, leading to a $T_{\rm hi}$ similar to the
$\approx 7$~MeV obtained for the neighboring bins in which multifragmentation 
represents the main reaction channel. 
Here, as the inspection of Fig.~\ref{fig:zbound} shows, heavy-residue 
production dominates but the fluctuations 
of $Z_{\rm max}$ are already large. For a possible interpretation, 
$T_{\rm hi}$ may, therefore, be assigned to 
multifragmentation events which are weak in this bin 
of $Z_{\rm bound}$ but perhaps exclusively selected 
by gating on the high-temperature $\alpha$ component.

The systematic errors associated with $T_{\rm hi}$ are enhanced by
the difficulty of reliably identifying the evaporation component in 
the spectra (Fig.~\ref{fig:alphas}). The quoted errors 
$\Delta y_{\rm hi} \approx 0.05$, obtained in the spectra-fitting procedure,
correspond to a temperature change of typically 
$\Delta T_{\rm hi} \approx 0.5$~MeV (Table~\ref{tab:thi}). 
The true uncertainties are likely to be 
larger even though 
secondary $\alpha$ components of 40\% in multifragmentation seem very reasonable.
For example, they compare well with 
the 26\% to 42\% determined with correlation techniques for central Xe on 
Sn collisions at 32 to 50~MeV per nucleon \cite{hudan03}. 

\begin{table}
\caption{\label{tab:thi}
He-Li double-isotope temperature $T_{\rm hi}$ derived by using only the 
high-temperature fraction $y_{\rm hi}$ of $\alpha$ particles 
(Fig.~\protect\ref{fig:alphas}), without further corrections, for four
bins of $Z_{\rm bound}$. $R_{\rm hi}$ is the corresponding effective correction factor. 
The errors given for $T_{\rm hi}$ and $R_{\rm hi}$ result
from the quoted uncertainty of $y_{\rm hi}$ obtained 
in the fitting procedure. 
}
\begin{ruledtabular}
\begin{tabular}{l c c c c}
$Z_{\rm bound}$ & $T_{\rm app}$ (MeV) & $y_{\rm hi}$ & $T_{\rm hi} (MeV) $ & $R_{\rm hi}$ \\
\hline
~0-20 & 8.0 & 0.62$\pm$.05 & 11.2$\pm$.8 & 1.40$\pm$.09 \\
20-40 & 5.5 & 0.57$\pm$.06 & ~7.2$\pm$.4 & 1.30$\pm$.07 \\
40-60 & 5.0 & 0.49$\pm$.04 & ~6.8$\pm$.3 & 1.37$\pm$.06 \\
60-80 & 4.5 & 0.28$\pm$.03 & ~7.9$\pm$.5 & 1.76$\pm$.11 \\
\end{tabular}
\end{ruledtabular}
\end{table}
 
This alternative and purely experimental method of considering the effects
of secondary decays may be seen as providing another estimate for the 
magnitude of systematic errors but also as globally confirming 
the obtained chemical breakup temperatures of 6 to 7~MeV for Au fragmentations.

Equilibrium including the kinetic degrees
of freedom is, apparently, not achieved in the present reactions. 
The slope temperatures of 12 to 17~MeV of the high-temperature component,
assigned here to multifragmentation events, are considerably higher than the 
chemical temperatures. They are, furthermore, nearly independent of the
fragment mass in the range $A \le 10$ \cite{odeh00} which rules out collective 
motions as their origin.
A significant collective flow should also become apparent in the impact-parameter
dependence of the kinetic energies which is not the case (Fig.~\ref{fig:alphas}).
Finite flow values have, nevertheless, been reported for spectator reactions 
but were derived from comparisons with statistical-model calculations (which 
assume kinetic equilibration) and by assigning a collective origin to measured 
excess energies \cite{avde02,lauret98,beau00}. An alternative explanation has 
been given by Odeh et al. \cite{odeh00} who have shown that, within the Goldhaber 
model of a random superposition of nucleon momenta \cite{gold74}, 
the larger slope 
temperatures are consistent with a chemical freeze-out at 6 to 8~MeV 
(see also Refs.~\cite{bauer95,gait00} for discussions of this scenario). 

\subsection{Chemical freeze-out}
\label{sec:chem}

The rise at small $Z_{\rm bound}$, i.e. high excitation energy, 
observed with $T_{\rm HeLi}$ (Figs.~\ref{fig:zbound},\ref{fig:target})
is well reproduced by most thermometers,
including $T_{\rm BeLi}$ which is 
derived from the $^{7,9}$Be and $^{6,8}$Li isotope ratios (Fig.~\ref{fig:qsm}).
There is, overall, good agreement between the different temperature
observables with the exception of those containing carbon isotopes.
In the latter cases, the apparent temperature values 
remain approximately constant with values between 4 and 5 MeV
(Fig.~\ref{fig:alltemp}).
For $T_{\rm CC}$ ($^{11,12}$C and $^{12,13}$C), a large
correction is required according to the QSM calculations 
(Fig.~\ref{fig:qsmcorr}) which lifts $T_{\rm CC}$ into the range of values 
assumed by the other temperature observables but $T_{\rm CLi}$ remains low. 
With the corrections according to Tsang, Xi et al. \cite{tsang97,xi99}, 
both temperatures involving carbon isotopes will remain near 5 MeV or below. 
In either case, a freeze-out state in chemical and thermal equilibrium including 
the carbon and possibly heavier isotopes seems to be excluded with a tendency
of heavier fragments having to 
come from colder regions of the system, as suggested by 
calculations within the quantum-molecular dynamics model 
\cite{goss97,zbiri07}.

A slightly different interpretation, based on measurements for a subset 
of the temperature observables studied here, 
has been proposed in Ref.~\cite{xi98}. 
For the central $^{86}$Kr on $^{93}$Nb collisions, 
two isotope thermometers $T_{\rm HeLi}$ and $T_{\rm CLi}$ were compared with 
the excited-state thermometers $^4$He, $^5$Li, and $^8$Be. Of these only the
$T_{\rm HeLi}$ temperature exhibited a rise with the bombarding energy which
suggested the conclusion that differences in the production environment of
$^3$He may be responsible for the singular behavior of $T_{\rm HeLi}$. 
While this cannot explain the common behavior of $T_{\rm BeLi}$ and 
$T_{\rm HeLi}$ observed in this work, it also points toward a more general 
scenario of different freeze-out conditions for 
different products, possibly also connected with the cooling prior to and during
the freeze-out process~\cite{liu06}. 

In contrast to central collisions of symmetric systems, there 
is no strong indication of collective radial flow in spectator 
decays, neither in the mass dependence of the kinetic energies of 
particles and fragments \cite{xi97,odeh00} nor in their absolute 
magnitude \cite{kunde95}, as discussed in the previous subsection. 
Therefore, if the different behavior of the excited-state and isotope 
temperatures for light fragments is assumed to have a common cause in both types of 
reactions, the collective flow by itself is not a very likely candidate for it. 
A more general common property of the two types of reactions is the 
disintegration of the whole system into many fragments and particles 
as the collisions become more violent.
The maximum multiplicities of intermediate-mass fragments,
after normalization with respect to the mass of the system, are of 
similar magnitude in central $^{197}$Au on $^{197}$Au collisions 
at 100 MeV per nucleon and in the spectator decays 
discussed here \cite{kunde95}.

\begin{figure}		
     \epsfysize=7.0cm
     \centerline{\epsffile{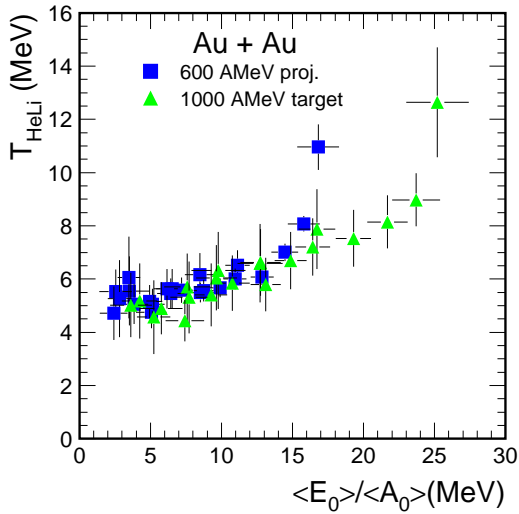}}
\caption{\label{fig:cc1} (Color online)
Caloric curves obtained for $^{197}$Au on $^{197}$Au at 600 and 1000 MeV 
per nucleon with temperatures $T_{\rm HeLi}$ and excitation energies from
calorimetry including the LAND data for neutrons \protect\cite{poch95,odeh99}.
}
\end{figure}

A rise of the temperature with decreasing $Z_{\rm bound}$ can be 
expected because the collisions become more violent. 
The fragment mass spectra change rapidly 
and become increasingly steep \cite{kreutz93}. Primarily, this indicates an
increase in deposited excitation energy \cite{ogilvie91,tamain06} which, however, 
may be associated with a rise in temperature.
The results obtained with the present data are shown in 
Fig.~\ref{fig:cc1}. $T_{\rm HeLi}$ is chosen as the temperature observable. 
The excitation energy was determined by calorimetry, as described in 
Refs.~\cite{poch95,odeh99}, considering in particular also the neutron 
energies and multiplicities measured with the LAND neutron detector
\cite{blaich92}.

The obtained caloric curves exhibit the overall behavior known from earlier
analyses~\cite{poch95} but the range of deposited energies 
$\langle E_0\rangle/\langle A_0\rangle$ covered at 1000 MeV
per nucleon incident energy extends to much higher values than at 600 MeV per 
nucleon which is inconsistent with the universality of the spectator decay 
that so clearly appears in other variables \cite{schuett96}. 
Technically, the difference can be traced back to the higher kinetic 
energies of neutrons in the frame of the projectile spectator at small
$Z_{\rm bound}$ and 1000 MeV per nucleon. The neutron energies measured 
with LAND were taken as representative also for the proton kinetic energies 
in this analysis. Preequilibrium nucleons from earlier reaction stages, apparently,
contribute at the highest excitation energies which, therefore, 
have to be considered as upper limits. 

\begin{figure}		
     \epsfysize=7.0cm
     \centerline{\epsffile{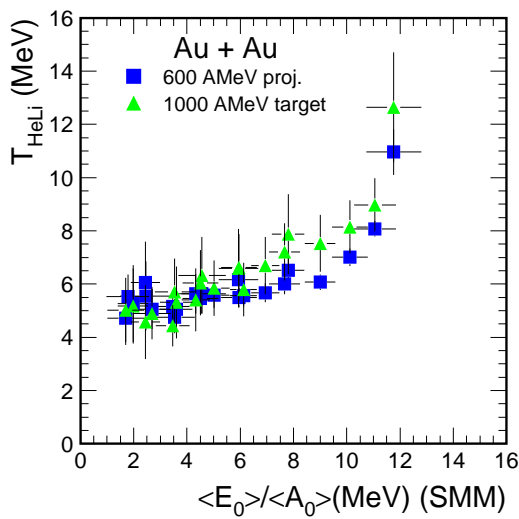}}
\caption{\label{fig:cc2} (Color online)
Caloric curves obtained for $^{197}$Au on $^{197}$Au at 600 and 1000 MeV 
per nucleon with temperatures $T_{\rm HeLi}$ and excitation energies as
obtained from the statistical-model description of the fragment observables 
\protect\cite{odeh00,odeh99}.
}
\end{figure}

Lower excitation energies, consistent with the invariance of the fragment decays with 
respect to the incident energy, were obtained from the analysis with the 
Statistical Multifragmentation Model with input parameters derived 
from the fragment observables alone \cite{xi97}. In this
so-called backtracing procedure (see also Refs.~\cite{botv95,desesq}),
an ensemble of input sources 
with varying mass and excitation energy is chosen and adjusted to 
reproduce the measured fragment-charge spectra and fragment correlations. 
The excitation energies of these sources are used in Fig.~\ref{fig:cc2}.
They should be considered as lower limits, however,
since the fragment kinetic energies are underestimated by the 
model (Refs.~\cite{schuett96,odeh00} and previous subsection). 

The significance and interpretations of the caloric curve for chemical freeze-out, 
widely discussed in the literature, have very recently been summarized 
by Keli{\'c} et al. \cite{kelic06}. Analyses with statistical and 
dynamical models show that the plateau-like part of the caloric curve 
may be identified with the transition region between the liquid nuclear 
phase and a disordered phase of particles and fragments (see, e.g., 
Refs.~\cite{dasg98,de97,lee97,schnack97,suga99,lefevre99,raduta00,furu04}). 
The methods and limitations of determining the excitation energy at the 
breakup stage have been reviewed by Viola and Bougault~\cite{violaboug06}.

The model comparison of the $Z_{\rm bound}$ dependence of the measured 
temperature avoids the problems of calorimetry.
In the backtracing analyses with the Statistical Multifragmentation Model presented 
in Refs.~\cite{xi97,bond98}, the observed small rise of the temperature 
with decreasing $Z_{\rm bound}$ has been quantitatively reproduced. 
The obtained apparent $T_{\rm HeLi}$, calculated with adjusted ensembles of 
fragmenting sources, was found to agree well with the experimental data. It was, 
furthermore, found to be slightly lower than the thermodynamical model
temperature, by an amount consistent with a
side-feeding correction of, on average, about 20\%. 
In more detail, the deviations turned out to be 
larger at the smaller excitations than at higher excitations
corresponding to small values of $Z_{\rm bound}$ (Fig.~7 in Ref.~\cite{xi97}
and Fig.~2 in Ref.~\cite{bond98}).
Within this model, the predicted plateau of the breakup temperature
$T \approx 6$~MeV for excitation energies between 4 and about 8 MeV is thus
consistent with the observed smooth rise of the chemical temperatures of light 
fragments in this interval.

In these calculations, the channels with smaller $Z_{\rm bound}$ are populated
by the decay of spectator sources of smaller mass and higher 
excitation, in accordance with the geometrical participant-spectator scenario. 
The rise of
the breakup temperature with decreasing $Z_{\rm bound}$, in this way, translates
into a mass dependence which is found to be rather generally obeyed in 
multifragment decays \cite{nato02}. In data from several experiments,
temperatures $T \approx 6$~MeV are 
associated with heavy systems of mass $A\approx200$ while higher 
temperatures up to $T \approx 8$~MeV are observed for lighter systems with 
$A<100$. The steady rise of the temperature with excitation energy 
for the very small systems has even been interpreted as indicating a breakup at
conditions close to the critical point \cite{ma04}.  

This systematic dependence has been associated with the concept 
of the limiting temperature at which excited homogeneous systems 
become unbound \cite{besp89}.  
According to these finite-temperature 
Hartree-Fock calculations, the limiting temperature should not only depend on the
mass of the system but, because of the changing Coulomb pressure, also  
on its isotopic composition. This prediction, however, has so far not
been confirmed experimentally. The $T_{\rm HeLi}$ temperatures for
target spectator decays initiated by $^{12}$C projectiles of 300 and 600 MeV
per nucleon on $^{112}$Sn and $^{124}$Sn targets differ by not more than 
0.5 MeV for similarly violent collisions \cite{lefevre05}, 
while a difference of about 1.5 MeV would be expected according to 
the calculations \cite{besp89}. However, in accordance with the predictions, 
the more proton-rich $^{112}$Sn system exhibits the lower temperature 
on average. 

\subsection{Internal excitation}
\label{sec:intex}

The measured excited-state temperatures of 5 to 6 MeV reflecting the 
internal excitation of the observed He, Li and Be isotopes are in good 
agreement with other measurements performed with the same method and
for reaction channels characterized by multiple fragment 
production~\cite{poch87,serf98,xi98,zhu95}.
Higher temperatures of 7 to 9 MeV have been 
reported for excited $^{6}$Li fragments 
in $^{14}$N on $^{197}$Au reactions at 35 MeV per nucleon after 
selecting the d-$\alpha$ pairs with the highest pair 
momenta while typical temperatures were 3 to 5 MeV also for this 
reaction~\cite{chit86}. Lower temperature values measured in compound 
reactions were found to be of the expected magnitude which confirms 
the sensitivity of the excited-state populations to the thermal 
characteristics of the reaction 
process \cite{schwarz93,nayak,dabr90,zhu94}. 

The internal excitation of heavier fragments at freeze-out in reactions at 
intermediate energies has been studied by measuring the correlations between light 
particles and fragments \cite{hudan03,marie98,staszel01}. For the reaction
$^{124}$Xe on Sn at 50 MeV per nucleon, the excitation energies of fragments over 
a wide range of $Z$ have been found to saturate at $E/A \approx 3$~MeV
\cite{hudan03,marie98} while a slightly lower value of 2.5 MeV has been reported 
for central $^{84}$Kr on $^{93}$Nb at 45 MeV \cite{staszel01} 
(for a review, see Ref.~\cite{verde06}). 
The corresponding temperatures of not more than 
5 MeV compare well with the present results and confirm their universal 
character as representing the internal excitation in fragment emissions.
Translated into time scales of evaporation processes, temperatures between 
4 and 6 MeV correspond rather generally to nuclear life times of the order 
of 50 to 200 fm/c \cite{friedlynch83,borderie92,suraud06} which cover
the range of time scales deduced for fragmentation 
reactions~\cite{viola99,schwarz01,rod02,viola06}.

These experimental methods probe the excitation of the fragments when they 
are sufficiently separated from the rest of the system, so that the correlations
of the decay products 
are not disturbed by further collisions. Similarly, the particle-unstable 
resonances will have to acquire their asymptotic quantum structure 
before they can be recognized. This is not likely to occur as long as they are 
connected to the nuclear medium \cite{ono04,botv06a,dani92,alm95}. 
The typical widths of about 1 MeV define very similar time scales 
of the order of 200 fm/c for their formation and decay.  
The obtained temperatures thus refer to a late stage in the reaction process 
whereas the chemical temperatures may reflect 
properties of earlier stages at which the fragments are still in contact
and their chemical compositions are still being established. 
In a disintegration scenario 
associated with cooling, an observed difference of these temperatures does not 
necessarily rule out that the two degrees of freedom are in equilibrium  
up to the time when one of them freezes out (cf. Refs.~\cite{poch87,chen87}).
On the other hand, molecular dynamics calculations consistently predict
differences between the internal product temperatures and those of the 
surrounding environment. 
According to quantum molecular dynamics calculations 
for the $^{197}$Au on $^{197}$Au reaction between 60 and 150 MeV per nucleon,
the relative motion of the constituents of asymptotically  
identified light fragments, e.g. $Z = 3$, corresponds to a nearly constant 
$T = 5$~MeV during most of the reaction time up to 200 fm/c \cite{zbiri07}. 
The environment, at the 
same time, cools from about 15 MeV to temperatures much below 5 MeV.
Also in classical molecular dynamics calculations, 
the internal temperatures evolve differently 
from those of the hot environment \cite{barz96,campi03}. 

These arguments may be sufficient to conclude that differences between 
the temperatures representing the chemical composition of the fragments 
and the internal excitation after their release have to be expected. 
The observed saturation of
the latter may, in fact, represent another and rather obvious evidence for
the existence of a limit up to which a nucleus can be 
excited \cite{koonin87}. In the case of surface emissions
from hot residues, this limit is evidently not yet reached by the system itself
which then can act as the heat reservoir determining both, the composition 
and internal excitation of the fragments. 
When the system becomes unstable with respect to bulk disintegration 
at higher excitations, composite products will still have to obey this limit 
in order to be recognized as such entities in the final partitions.

It is, nevertheless, a rather intriguing result that internal temperatures of
5 MeV are universally observed in the fragmentation channels throughout 
the wide range of reactions covered by the many experiments, including, e.g., 
also the measurement of bremsstrahlung photons~\cite{enter01}.
The question arises whether this would have to be directly connected to 
the process of fragment formation in order to exhibit this generality.
The problems associated with predicting internal fragment excitations at
freeze-out in an expansion scenario have been recognized early on 
\cite{boal89} after the first data on excited-state temperatures had 
become available \cite{morri84,poch85}. Values near $T = 5$~MeV have been 
very consistently obtained even though remarkably different theoretical 
approaches were followed. 
The presented reasons are numerous and of different origin as,
e.g., coalescence heating \cite{fried88,barz89} or the onset of instabilities
\cite{fuchs97}. 

A mechanism leading to invariant internal excitations is also constituted 
by the Goldhaber model \cite{gold74} which has been found to provide a basis for the
understanding of the fragment kinetic energies in reactions at relativistic
and also intermediate energies \cite{gait00,odeh00,bauer95,luka02}. 
As an example, the clusterization mechanism proposed  
for the production of fragments of mass number $A$ in peripheral collisions 
at intermediate energies involves a random picking of $A$ nucleons from a 
cold Fermi sphere \cite{luka03}. Their momenta in the center-of-mass system 
of the formed fragment correspond to those of an excited system.  
This has been tested by fitting finite-temperature Fermi distributions 
to the mean momentum distributions obtained with a Monte Carlo procedure 
following this prescription. The resulting temperatures drop, approximately
like $1/\sqrt{A}$, from about 7 MeV for $A=5$ to smaller values for larger 
clusters \cite{luka03}. For intermediate-mass fragments, this is of the order
of magnitude that is observed. 
More importantly, this mechanism is fairly independent of 
the global course of the reaction and of the excitation of the emitting system.

\section{Conclusion and outlook}
\label{sec:conc}

A summary has been presented of a series of temperature measurements for
spectator decays following the excitation of $^{197}$Au nuclei in relativistic
collisions at energies up to 1 GeV per nucleon. The methods used permit the
determination of the temperature representing the assumed thermal population of
the considered degrees of freedom. They are model independent except for the
assessment of the effects of sequential decay which, however, are important,
in particular for temperatures derived from double isotope-yield ratios.
Kinetic energies and the kinetic temperatures associated with them were only
briefly addressed. 
Rather than collective motion, the Fermi motion of nucleons in the colliding nuclei 
are evidently responsible for their high values which considerably 
exceed the thermal and chemical breakup temperatures in the reactions studied here. 

The main observation derived from the data and discussed in detail is the 
divergence of the isotope and excited-state temperatures with increasing 
excitation of the spectator system. The chemical and thermal freeze-outs 
cease to coincide if bulk 
disintegrations of major parts of the excited intermediate system start 
to become important, i.e., at the point of transition from residue formation 
to multifragment production. The higher values of temperatures derived from 
double-isotope ratios suggest that the chemical compositions are established 
prior to the thermal freeze-out stage when the products finally disconnect from 
the system. The thermal freeze-out temperatures deduced from excited-state
populations of particle-unstable resonances saturate around $T$ = 5 MeV, a value
common to several classes of fragmentation reactions. In particular,
the patterns observed in 
explosive fragmentations following central collisions of heavy systems
and in the largely equilibrated spectator decays studied here are rather 
similar. 

The chemical freeze-out temperatures have characteristic properties in 
common with the partitions themselves as, e.g., the invariance properties of 
$Z_{\rm bound}$-sorted observables. Here, the invariance with respect to the mass
of the collision partner and the invariance with the bombarding energy in the
studied range of 600 to 1000 MeV per nucleon have been explicitly demonstrated.
As a new result, it is found that the $T_{\rm BeLi}$ thermometer based on
fragments with $Z \geq 3$ follows the results obtained with the frequently used
$T_{\rm HeLi}$ thermometer. Also the single ratios of isotope yields exhibit the 
$Z_{\rm bound}$ dependence that is expected from the binding-energy difference
in chemical equilibrium. Only the double ratios containing carbon
isotopes do not consistently support the general trend to higher temperatures at 
smaller $Z_{\rm bound}$. With the present data, it has not been possible to 
firmly resolve whether this is caused by nuclear-structure effects in particular 
combinations of isotopes or whether, more likely, it indicates some restriction 
in the validity of a concept of global chemical equilibrium at breakup. Heavier
products, unable to survive in very hot environments, can only come from cooler 
regions of the system.

The uncertainty of secondary-decay corrections remains the limiting 
factor in the quantitative analysis. The corrections are well under control at
the lower temperatures of up to about 4 MeV and, in particular, in the 
case of emissions from excited residue nuclei. 
The uncertainties increase with the 
temperature and approach the MeV order of magnitude once typical breakup 
temperatures of 6 MeV are reached or exceeded. 
This was illustrated by comparing several 
different correction methods. While it is highly desirable to improve this 
situation, it has, nevertheless, been possible to establish important
trends, in good agreement with results from other experiments.
A remaining open problem is constituted by the somewhat unclear role of 
limiting temperatures which seem to influence the internal fragment excitations
but not the dependence of the chemical temperatures on the isotopic composition
of the system, at least not in the predicted magnitude.

The authors would like to thank A.S. Botvina, J.B. Natowitz, W. Reisdorf, 
and M.B.~Tsang for fruitful discussions. 
This work was supported by the European Community under
contract ERBFMGECT950083.

\end{document}